\documentclass[longbibliography,aps,prl,twocolumn,superscriptaddress,showkeys,amsfonts,amssymb,amsmath,floatfix]{revtex4-2}
\usepackage{inputenc}
\usepackage{amsmath,amsbsy,amsfonts,revsymb}
\usepackage{graphicx}
\usepackage{bm}
\usepackage[dvipsnames]{xcolor}
\usepackage{chemformula}
\usepackage[separate-uncertainty=true,multi-part-units=single]{siunitx}
\usepackage{bm}
\usepackage{mathtools} 
\usepackage{placeins}
\usepackage{multirow}
\usepackage[colorlinks=true,linkcolor=blue,citecolor=blue,urlcolor=blue]{hyperref}
\usepackage{soul}

\makeatletter
\def\maketitle{
\@author@finish
\title@column\titleblock@produce
\suppressfloats[t]}
\makeatother

\DeclareSIUnit \mub {\ensuremath{\mu_{B}}}

\newcommand\mrm[1]{\mathrm{#1}}

\newcommand{\NCCVO}{NaCa$_{2}$Cu$_{2}$(VO$_{4}$)$_{3}$}
\newcommand{\NCMVO}{NaCa$_{2}$Mg$_{2}$(VO$_{4}$)$_{3}$}
\newcommand{\tCW}{\theta_{\mrm{CW}}}

\begin{document}

\title{Evidence for spin liquid behavior in the frustrated three-dimensional $\mathbf{S = 1/2}$ Heisenberg garnet NaCa$_{\mathbf{2}}$Cu$_{\mathbf{2}}$(VO$_{\mathbf{4}}$)$_{\mathbf{3}}$}

\author {Y. Alexanian}\email{yann.alexanian@unige.ch}
\altaffiliation[]{current address: Department of Quantum Matter Physics, University of Geneva, CH-1211 Geneva, Switzerland}
\affiliation{Institut Laue-Langevin, 71 avenue des Martyrs, CS 20156, 38042 Grenoble cedex 9, France}
\author{R. Kumar}
\affiliation{Université Paris-Saclay, CNRS, Laboratoire de Physique des Solides, 91405, Orsay, France}
\author{H. Zeroual}
\affiliation{Université Paris-Saclay, CNRS, Laboratoire de Physique des Solides, 91405, Orsay, France}
\author{B. Bernu}
\affiliation{Sorbonne Université, CNRS, Laboratoire de Physique Théorique de la Matière Condensée, 75252 Paris Cedex 05, France}
\author{L. Mangin-Thro}
\affiliation{Institut Laue-Langevin, 71 avenue des Martyrs, CS 20156, 38042 Grenoble cedex 9, France}
\author{J. R. Stewart}
\affiliation{ ISIS Neutron and Muon Source, Rutherford Appleton Laboratory, Didcot OX11 0QX, United Kingdom}
\author{J. M. Wilkinson}
\affiliation{ ISIS Neutron and Muon Source, Rutherford Appleton Laboratory, Didcot OX11 0QX, United Kingdom}
\author{S. Bhattacharya}
\affiliation{Université Paris-Saclay, CNRS, Laboratoire de Physique des Solides, 91405, Orsay, France}
\author{P. L. Paulose}
\affiliation{Department of Condensed Matter Physics and Materials Science, TIFR, Mumbai 400 005, India}
\author{F. Bert}
\affiliation{Université Paris-Saclay, CNRS, Laboratoire de Physique des Solides, 91405, Orsay, France}
\author{P. Mendels}
\affiliation{Université Paris-Saclay, CNRS, Laboratoire de Physique des Solides, 91405, Orsay, France}
\author{B. F\aa k}
\affiliation{Institut Laue-Langevin, 71 avenue des Martyrs, CS 20156, 38042 Grenoble cedex 9, France}
\author{E. Kermarrec}\email{edwin.kermarrec@universite-paris-saclay.fr}
\affiliation{Université Paris-Saclay, CNRS, Laboratoire de Physique des Solides, 91405, Orsay, France}

\date{\today}
	
\begin{abstract}
Three-dimensional quantum spin liquids have remained elusive, hindered by reduced quantum fluctuations from larger lattice connectivity inherent to high-dimensional systems. Here, we investigate the remarkable persistence of dynamical short-range magnetic correlations in the nearly body-centered cubic garnet \NCCVO\ down to $T = \SI{50}{\milli\kelvin}$, two orders of magnitude below its Curie-Weiss temperature. Using a combination of neutron and muon spectroscopies plus numerical simulations, we demonstrate that a dynamical regime emerges, characterized by a dual response in the inelastic spectrum composed of short-live dispersive excitations and a quasi-elastic component. Strongly frustrated exchange interactions combined with subtle temperature-dependent Jahn-Teller spin-lattice effects are a plausible mechanism to explain the origin of this spin-liquid behavior.
\end{abstract}

\maketitle

Quantum spin liquids (QSLs) are long-range entangled states of matter, that evade thermodynamic phase transitions and remain dynamically disordered due to quantum fluctuations \cite{Lacroix2011,Balents2010,Broholm2020}, yet with defining signatures such as topological order or fractionalization. While one-dimensional (1D) QSLs are well-established \cite{Haldane1983a,Haldane1983b,Schollwock2004,Giamarchi2004}, they appear in more constrained forms compared to the diverse theoretical predictions for higher dimensions. Over the past few decades, extensive research has focused on two-dimensional (2D) QSL candidates built on triangular lattice motifs \cite{Mendels2015,Li2020,Khuntia2020}. The mineral herbertsmithite ZnCu$_{3}$(OH)$_{6}$Cl$_{2}$ stands out as the most promising realization of the $S=1/2$ Kagome Heisenberg antiferromagnet exhibiting several defining characteristics of QSLs \cite{Mendels2007,Han2012}. Conversely, three-dimensional (3D) QSL candidates have long been scarce, since the higher lattice connectivity limits the degeneracy of low energy states necessary for QSL formation. The recent growing variety of experimentally discovered low-connectivity three-dimensional (3D) frustrated lattices (hyperkagome \cite{Okamoto2007,Khatua2022}, pyrochlore \cite{Subramanian1983,Gardner2010}, trillium \cite{Zivkovi2021,Bulled2022,Boya2022}, etc.) opens new perspectives on this long-standing challenge \cite{Moessner1998,Canals1998,Hopkinson2006,Hopkinson2007,Savary2017,Li2024}. Rare-earth-based pyrochlore oxides provide notable cases, with exotic ground states stabilized by strong anisotropic interactions and large magnetic moments \cite{Rau2019}. Nevertheless, the complex nature of $4f$ electron states and their small magnetic energy scale complicate the identification of the key mechanisms underlying their low-temperature behavior. While a consensus on the existence of a true 3D QSL has thus not yet been reached, Ce-based \cite{Smith2022, Poree2024} and charge disordered Tb-based pyrochlores currently show great promises \cite{Sibille2017,Alexanian2023}.

An alternative approach to realize a 3D QSL involves frustration arising from multiple competing exchange within seemingly simpler lattices. A few recent candidates, such as Ca$_{10}$Cr$_{7}$O$_{28}$ \cite{Balz2016,Balz2017} and PbCuTe$_{2}$O$_{6}$ \cite{Koteswararao2014,Khuntia2016,Hong2023}, illustrate how these competing interactions can stabilize unconventional magnetic states. Also very promising is the garnet family X$_{3}$Y$_{2}$(ZO$_{4}$)$_{3}$ where atoms at the $Y$-site form a body-centered cubic (bcc) lattice \cite{Bayer1965,Geller1967}. In Ca$_{3}$Cu$_{2}$GeV$_{2}$O$_{12}$, the Cu$^{2+}$ spins $S =1/2$, naturally prone to quantum fluctuations, realize the bcc lattice, preventing long-range magnetic ordering down to $T = \SI{0.35}{\kelvin}$ \cite{Lussier2019}. However, it has a low magnetic energy scale ($\tCW = -\SI{0.93}{\kelvin}$), and the role of paramagnetic impurities in determining its ground state remains unclear. 

Interestingly, several members of this magnetic garnet family were investigated long before the concept of QSLs had emerged \cite{Kazei1976,Kazei1980,Kazei1982,Shender1982,Kazei1983}. One notable example is the copper-based material \NCCVO, where Cu$^{2+}$ ions occupy a bcc lattice at room temperature and Na$^+$/Ca$^{2+}$ shared the same 24$c$ site, providing a natural source of chemical disorder. Below $T \approx \SI{250}{\kelvin}$, the crystallographic structure undergoes a small structural transition to a tetragonal phase ($c/a = 0.985$ at $\SI{4.2}{\kelvin}$), likely driven by cooperative Jahn-Teller effects \cite{Kazei1982}. Early macroscopic studies revealed magnetic correlations from $T = \SI{25}{\kelvin}$, but no long-range magnetic order was observed down to at least $T = \SI{0.2}{\kelvin}$ \cite{Kazei1983}. 

This is particularly intriguing in light of recent numerical studies of the  $J_1 - J_2 - J_3$ bcc Heisenberg model \cite{Ghosh2019,Sonnenschein2020}. No classical spin liquid phase is stabilized and it is only in the quantum $S=1/2$ limit that a small region of the ground-state phase diagram is found to be paramagnetic. The close proximity of \NCCVO\ to a perfect bcc lattice and the absence of magnetic order in this system thus warrant further investigations.

\begin{figure*}[ht!]
\includegraphics[width=1\textwidth]{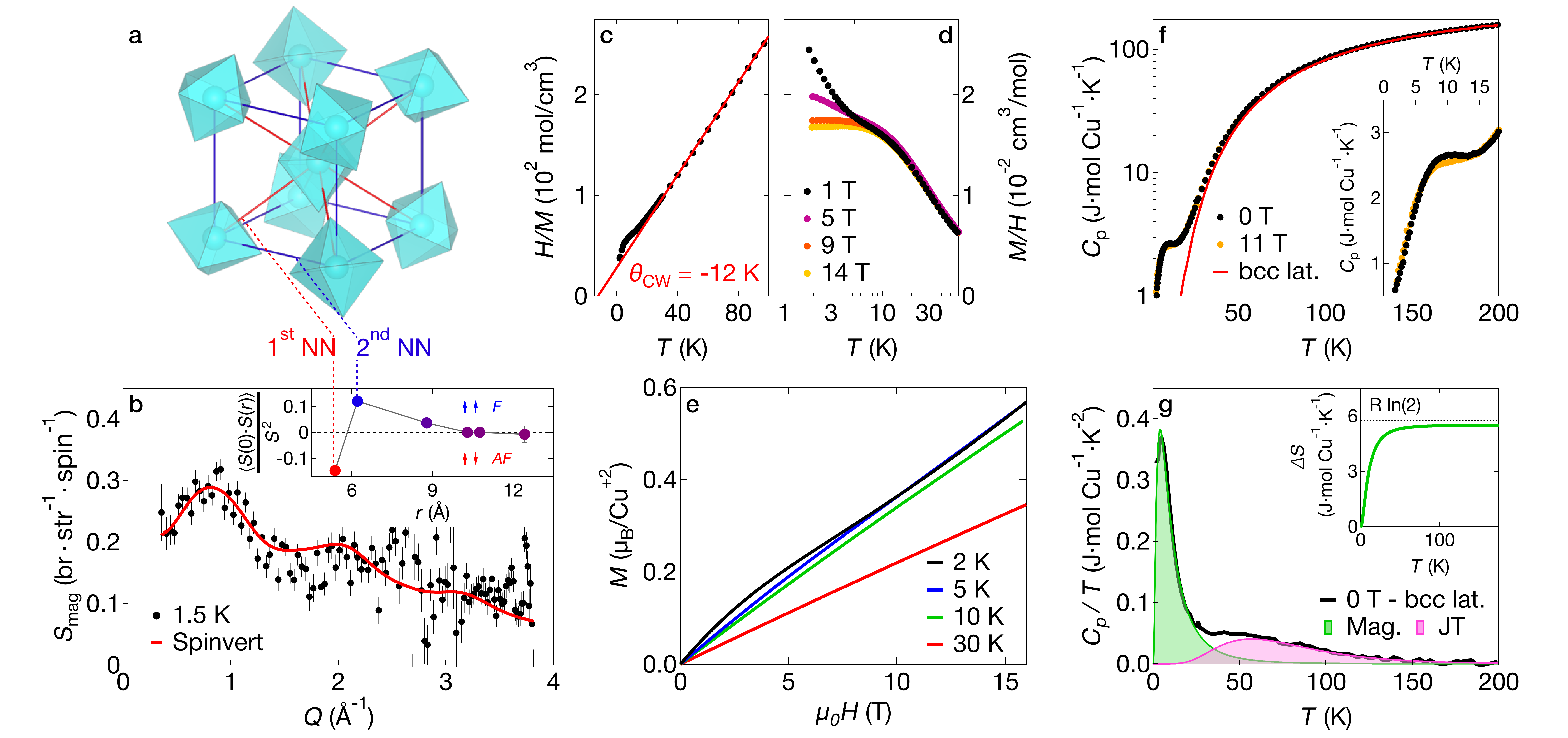 }
\caption{\textbf{a} Perfect bcc lattice of Cu$^{2+}$ ions in \NCCVO\ at room temperature. Oxygens O$^{2-}$ sits at the polyhedral vertices. First- and second nearest-neighbors are indicated by red and blue lines, respectively.  \textbf{b} Magnetic scattering intensity from spin-polarized measurements at $T=\SI{1.5}{\kelvin}$. The red line shows the \textsc{SPINVERT} refinement using isotropic spins and $2\times2\times2$ cubic unit cells. Residual contribution from nuclear Bragg peaks contributes above $\SI{2.5}{\angstrom^{-1}}$. Inset: normalized spin correlation function extracted from the refinement. \textbf{c} Inverse magnetic susceptibility with its high-temperature fit ($\SIrange{50}{250}{\kelvin}$). \textbf{d} Magnetic susceptibility measured at various magnetic field. \textbf{e} Magnetization up to $\mu_{0}H=\SI{16}{\tesla}$ at different temperatures. \textbf{f} Temperature dependence of the specific heat at $\SI{0}{\tesla}$ and $\SI{11}{\tesla}$. The bcc lattice (bcc lat.) contribution is obtained from measurements of the non-magnetic analogue \NCMVO. \textbf{g} Zero-field specific heat of \NCCVO\ subtracted from the bcc lattice contribution (black line), and fits of the magnetic (Mag., green area) and Jahn-Teller (JT, pink area) contributions. Inset: temperature variation of the magnetic entropy, obtained from the integration of the magnetic specific heat.}
\label{fig:1}
\end{figure*}

In this study, we demonstrate that the quantum spins of \NCCVO\ remain dynamic down to at least $T=\SI{0.05}{\kelvin}$. Our data further reveal temperature-dependent Jahn-Teller effects along with a subtle interplay of quasi-elastic and inelastic dynamics. Based on a numerical analysis, we attribute this characteristic to strong magnetic frustration among incipient ferromagnetic spin chains mediated by spin-lattice effects. These competing interactions may ultimately promote the unconventional three-dimensional spin liquid behavior of \NCCVO.


\section{Results}

\subsection{Polarized neutron diffraction and bulk characterisation : evidences for magnetic correlations}
\vspace{-0.4cm}

In Fig. \ref{fig:1}, we present experimental evidences that \NCCVO\ exhibits a correlated magnetic state without long range order. Our polycrystalline sample was synthesized via solid-state reaction, and its crystallographic structure was checked by X-ray diffraction at $T=\SI{298}{\kelvin}$ (see Fig.~\hyperref[fig:1]{\ref*{fig:1}a} and Supplementary Material (SM) Sec. A \cite{SM}). 

Magnetic correlations are first revealed through our polarized neutron diffraction measurements at $T=\SI{1.5}{\kelvin}$ recorded on D7 at ILL \cite{D7, ILLdata} (Fig.~\hyperref[fig:1]{\ref*{fig:1}b}). No magnetic Bragg peaks are observed, but oscillations in the magnetic scattering function $S_{\rm mag}$ with a prominent maximum at $Q=\SI{0.9}{\angstrom^{-1}}$ are evident. This establishes the absence of long-range order but the presence of magnetic correlations, no longer visible at $T=\SI{300}{\kelvin}$ (see SM Sec. B \cite{SM}). We performed Monte Carlo simulations using the \textsc{SPINVERT} software \cite{Paddison2013} with $N\times N\times N$ cubic unit cells of isotropic spins to characterize the correlations. Good agreement was achieved for $N=2$ (without improvement when further increasing $N$) while models restricted to Ising or XY spins did not reproduce the observed data. We found antiferromagnetic spin correlations $\left\langle \bm{S}(0) \cdot \bm{S}(r) \right\rangle/S^2$ for first nearest-neighbors ($-0.15$) and ferromagnetic ones for second nearest-neighbors ($0.12$), located respectively at $\SI{5.38}{\angstrom}$ and $\SI{6.21}{\angstrom}$, and rapidly decreasing beyond these distances. Introducing the small tetragonal distortion gave similar results, as expected in view of the minute atomic displacement of $\sim\SI{0.1}{\angstrom}$. The magnetic correlations obtained from our powder sample data correspond to the averaged correlations in the three different axis directions. The expected magnetic scattering from single crystals calculated with the \textsc{SCATTY} software \cite{Paddison2019} is shown in SM Sec. B \cite{SM}. 


Signatures of magnetic interactions affecting the system at much higher temperatures - and hence  magnetic frustration - are evident in our inverse magnetic susceptibility data shown in Fig.~\hyperref[fig:1]{\ref*{fig:1}c}. The curve follows a Curie-Weiss law down to $T=\SI{30}{\kelvin}$, below which it begins to deviate from linearity. Fitting the high-temperature range above $\SI{50}{\kelvin}$ yields a Curie-Weiss temperature of $\theta_{\mrm{CW}} = -\SI{12}{\kelvin}$, indicating dominant antiferromagnetic interactions. Notably, another change of slope is observed at lower temperatures, between $T=\SIrange{5}{10}{\kelvin}$. As shown in Fig.~\hyperref[fig:1]{\ref*{fig:1}d}, the magnetic susceptibility becomes field-dependent at such temperatures, saturating only above $\SI{9}{\tesla}$. Subtracting the high-field from low-field data rules out a contribution from paramagnetic impurities, as the result does not follow a Curie law. Further insight comes from our magnetization measurements up to $\mu_{0}H=\SI{16}{\tesla}$ accross temperatures from $T=\SI{2}{\kelvin}$ to $T=\SI{30}{\kelvin}$ (Fig.~\hyperref[fig:1]{\ref*{fig:1}e}). The magnetization evolves smoothly at each temperature, and reaches $\SI{0.56}{\mu_{\mrm{B}}}$ per Cu$^{2+}$ ion at $\SI{2}{\kelvin}$ and $\SI{16}{\tesla}$, lower than the expected $gS\mu_{\mrm{B}} = \SI{1}{\mu_{\mrm{B}}}$ and far from saturation. Additionally, a distinct feature emerges at low magnetic field for $T=\SI{2}{\kelvin}$, with a significantly faster polarization of some of the spins up to 6 T, as compared to the $T=\SI{5}{\kelvin}$ data. This behavior correlates with the magnetic field dependence of the susceptibility below $T=\SI{5}{\kelvin}$ and indicates the presence of a second magnetic energy scale of ferromagnetic sign.

Specific heat data are presented in Fig.~\hyperref[fig:1]{\ref*{fig:1}f}. A clear departure from the bcc lattice contribution (derived from measurements on the non-magnetic analogue \NCMVO, see Ref. \cite{Hardy2003} for the method) is evident below $T=\SI{50}{\kelvin}$: this is another indication of magnetic correlations. Consistent with the absence of saturation in our magnetization data, only a marginal response to a magnetic field of $\mu_{0}H=\SI{11}{\tesla}$ is observed. The specific heat subtracted from the bcc lattice signal and divided by the temperature (Fig.~\hyperref[fig:1]{\ref*{fig:1}g}) reveals a pronounced peak around $\SI{5}{\kelvin}$ with a long tail extending up to $T=\SI{200}{\kelvin}$. Since signatures of magnetic correlations are only evident below $T=\SI{50}{\kelvin}$ in both the specific heat and magnetic susceptibility data, we attribute the long tail to another lattice effect. This is not surprising since \NCCVO\ is Jahn-Teller active below $T=\SI{250}{\kelvin}$, unlike the perfectly bcc non-magnetic analogue \NCMVO\ from which we extracted the initial lattice contribution. To estimate the magnetic specific heat, we  fit the magnetic and Jahn-Teller contributions with two asymmetric functions. The integration of the magnetic signal reaches $\SI{95}{\percent}$ of $R\ln(2)$ at $T=\SI{50}{\kelvin}$. This simple, phenomenological, analysis suggests that nearly all the expected magnetic entropy is recovered, although one cannot exclude a $\sim \SI{10}{\percent}$ missing entropy and therefore a potential partial magnetic ordering at even lower temperatures (see SM Sec. C \cite{SM} for details).

\subsection{Muon spin relaxation: a dynamical state down to 50~mK}       
\vspace{-0.4cm}

\begin{figure*}
\includegraphics[width=1\textwidth]{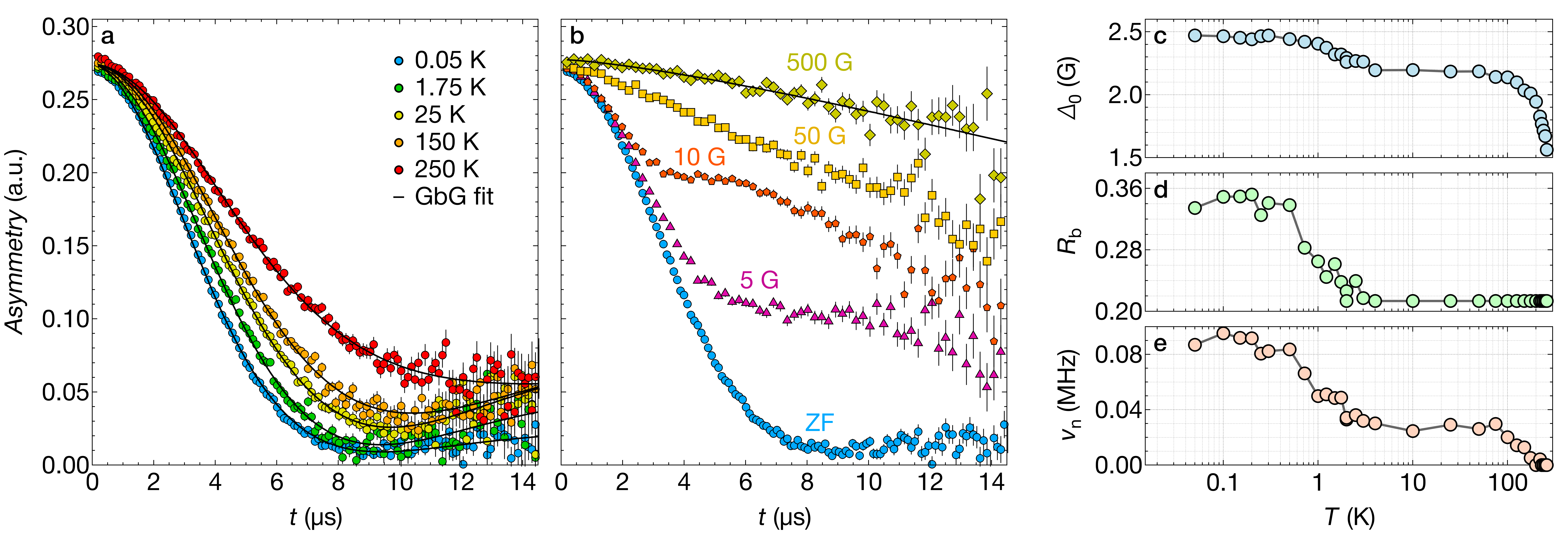 }
\caption{\textbf{a} Temperature dependence of the zero-field relaxation. Lines are fits to a dynamical Gaussian-broadened-Gaussian model (GbG) (see text).  
\textbf{b} Relaxation measured under longitudinal applied fields, from 0 (ZF) to 500 G, at $T = 50$~mK. Under 500~G, the residual relaxation reflects fast spin dynamics. Black line is a stretched exponential fit.  
\textbf{c}, \textbf{d}, \textbf{e} Evolution in temperature of the extracted fitting parameters $\Delta_0$, $R_{\rm b}$ and $\nu_{\rm n}$ (see Eq. \ref{GbG}).
}
\label{fig:2}
\end{figure*}

To confirm the absence of long-range order or any other spin freezing at lower temperatures, we performed muon spin relaxation ($\mu$SR) down to $\SI{0.05}{\kelvin}$ (Fig. \ref{fig:2}), on the MuSR spectrometer at ISIS \cite{muSRdata} (see SM Sec. D \cite{SM} for technical details). The temperature evolution of a selection of zero-field relaxation from $\SI{300}{\kelvin}$ down to $\SI{0.05}{\kelvin}$ is shown in Fig.~\hyperref[fig:2]{\ref*{fig:2}a}. At $\SI{300}{\kelvin}$, the Cu$^{2+}$ moments are in a fast fluctuating regime and the Gaussian shape of the relaxation at early times reflects the nuclear static field.
At $\SI{50}{\milli\kelvin}$, longitudinal field measurements quickly recover almost the full asymmetry, confirming the nuclear static origin of the Gaussian relaxation. Above 50~G, a slowly exponentially decaying relaxation points to the presence of fast, persistent, spin dynamics of electronic origin (Fig.~\hyperref[fig:2]{\ref*{fig:2}b}). A first estimate of the fluctuation frequency $\nu_{\rm e}$ can be derived from a stretched exponential fit of the relaxation, $Ae^{-(\lambda t)^{\beta}}$.  Such fit of the 500~G data leads to $\lambda = 0.024(1)$~$\mu$s$^{-1}$ and $\beta = 1.38(6)$, which, assuming a fluctuating moment for a Cu$^{2+}$ ion of 1~$\mu_B$, gives $\nu_{\rm e} \simeq 2\gamma_{\mu}^2\Delta^2 / \lambda = 44(2)$~GHz using the Redfield formula, with $\Delta$ the fluctuating field and $\gamma_{\mu} = 851.4$~Mrad.T$^{-1}$ the muon gyromagnetic ratio. This confirms the fast fluctuation regime with $\nu_{\rm e} \gg \gamma_{\mu}\Delta \simeq 23$~MHz. 
Our $\mu$SR data shows that the Cu$^{2+}$ moments remain dynamic down to $\SI{50}{\milli\kelvin}$, qualifying the garnet \NCCVO\ as a potential spin liquid candidate.

We now turn to the evolution of the zero-field relaxation over the whole $\SIrange{0.05}{300}{\kelvin}$ range. The absence of a sharp minimum prevents to use the conventional Kubo-Toyabe function, and we found the best modelization of the data using the dynamical Gaussian-broadened-Gaussian Kubo Toyabe function \cite{Dalmas2011, Noakes1997}. The pronounced Gaussian shape coupled to the singularly reduced local dip near $\SIrange{8}{10}{\micro\second}$ strongly suggest to use the quasi-static approximation with the following expression \cite{Dalmas2011}:
\begin{equation}
\begin{gathered}
P_{\rm GbG} (t) = \frac{1}{3} \exp \left( -\frac{2}{3} \nu_{\rm n} t \right) \\
+ \frac{2}{3} \left( \frac{1 }{1  + R_{\rm b}^2 \gamma_{\mu}^2\Delta_{0}^2 t^2} \right)^{3/2}
\left( 1 - \frac{\gamma_{\mu}^2\Delta_{0}^2 t^2}{1  + R_{\rm b}^2 \gamma_{\mu}^2\Delta_{0}^2 t^2} \right)\\
\times \exp \left[ - \frac{\gamma_{\mu}^2\Delta_{0}^2 t^2}{2 ( 1 + R_{\rm b}^2 \gamma_{\mu}^2\Delta_{0}^2 t^2)} \right]
\end{gathered}
\label{GbG}
\end{equation}

This function models a relaxation due to a homogeneous disorder, yet within a disordered static scenario of a collection of Gaussian field distributions. This results in a three parameters model: the width of the static field Gaussian distribution $\Delta_0$, the ratio $R_{\rm b} = \Delta_{\mrm{G}} / \Delta_0$  where $\Delta_{\mrm{G}}$ is the standard deviation of the Gaussian-weighted initial distribution, and the fluctuation frequency $\nu_{\rm n}$ characterizing the dynamics.  


This successfully captures the behavior of the relaxation over the whole temperature range, and the dependence of the three parameters are shown in Figs.~\hyperref[fig:2]{\ref*{fig:2}c}, \hyperref[fig:2]{\ref*{fig:2}d} and \hyperref[fig:2]{\ref*{fig:2}e}. From $\SI{300}{\kelvin}$ to $\SI{50}{\kelvin}$, the distribution width $\Delta_0$ due to nuclear fields increases with temperature up to 2.2~G. This is somewhat unusual as this value depends on the structural positions of neighboring atoms. Here, this naturally reflects the smooth evolution of the structural distortion between $\num{250}$ and $\SI{50}{\kelvin}$, as reported by x-ray diffraction \cite{Kazei1982}. Thus, our muon data appear to be strongly sensitive to the Jahn-Teller driven structural transition below $\SI{250}{\kelvin}$ that clearly modifies the local environnement of the muon, and then the value of $\Delta_0$. Interestingly, $\Delta_0$ is found to increase again just below $\SI{10}{\kelvin}$, where the magnetic correlations set in, which hints at a subtle atomic and/or orbital modification. The parameter $R_{\rm b}$ essentially shares the same temperature dependence as $\Delta_0$ below $\SI{10}{\kelvin}$, with values ranging from $\sim 0.2$ to $0.35$. This indicates a moderate structural disorder, likely caused by the Na/Ca double occupation of the same cristallographic site (Wyckoff 24\textit{c}) and Jahn-Teller vibrational modes. This provides a natural explanation for the Gaussian-broadened-Gaussian model which captures the multiple muon environments.   
Finally, the parameter $\nu_{\rm n}$ is important as it directly tracks the dynamics probed by the muon. Although $\nu_{\rm n}$ and $\Delta_0$ seem highly correlated, fixing one of them considerably degrades the quality of the fit, and thus we conclude to the physical relevance of both parameters. At first, $\nu_{\rm n}$ slightly increases upon cooling from $\SI{300}{\kelvin}$ and levels off below $\sim\SI{50}{\kelvin}$. Similarly to $\Delta_0$, a second increase occurs below $\SI{10}{\kelvin}$ and then $\nu_{\rm n}$ remains constant below $\SI{0.5}{\kelvin}$. In our GbG model, the muon mainly captures the fluctuations of the nuclear fields at the frequency $\nu_{\rm n}$. The rise of both $\nu_{\rm n}$ and $\Delta_0$ below $\SI{10}{\kelvin}$ signals subtle structural modifications, possibly driven by the onset of magnetic correlation when $k_{\mrm B} T \sim J$. 
We note the tiny anomaly around $\SI{0.2}{\kelvin}$ for all parameters, reminiscent of the observation of a kink in magnetic susceptibility \cite{Kazei1983} and interpreted as indication of magnetic order. Our muon data, sensitive to both structural change and magnetism, rather shows a minor effect which does not lead to any major spin freezing or ordering.
Although the determination of the muon position would be required to fully discard the hypothesis of a cancellation of the static magnetic field at the muon site, we considered it to be very unlikely in the present case. Indeed, given the large unit cell with many different oxygen positions, such cancellation would not occur for all muon sites. In contrast, the minute change of relaxation observed at $\SI{0.2}{\kelvin}$ is detected for the whole asymmetry and thus is probed by all the muon sites. A tiny magnetic impurity could be held responsible for the anomaly in the magnetic susceptibility. Our muon data would be compatible with such hypothesis, provided the impurity level does not exceed $\sim \SI{0.3}{\percent}$ of the sample volume. 

In summary, our $\mu$SR results support the presence of a correlated regime below $\SI{10}{\kelvin}$, in agreement with polarized neutron diffraction, yet with Cu$^{2+}$ moments which remain fully dynamical down to $\SI{50}{\milli\kelvin}$. Furthermore, it reveals the existence of two types of excitations with different timescales: (i) a slow dynamic ascribed to the fluctuations of nuclear fields and linked to the Jahn-Teller vibrations, and (ii) fast fluctuations relevant to the Cu$^{2+}$ electronic moments.

\begin{figure}[t!]
\includegraphics[width=1\columnwidth]{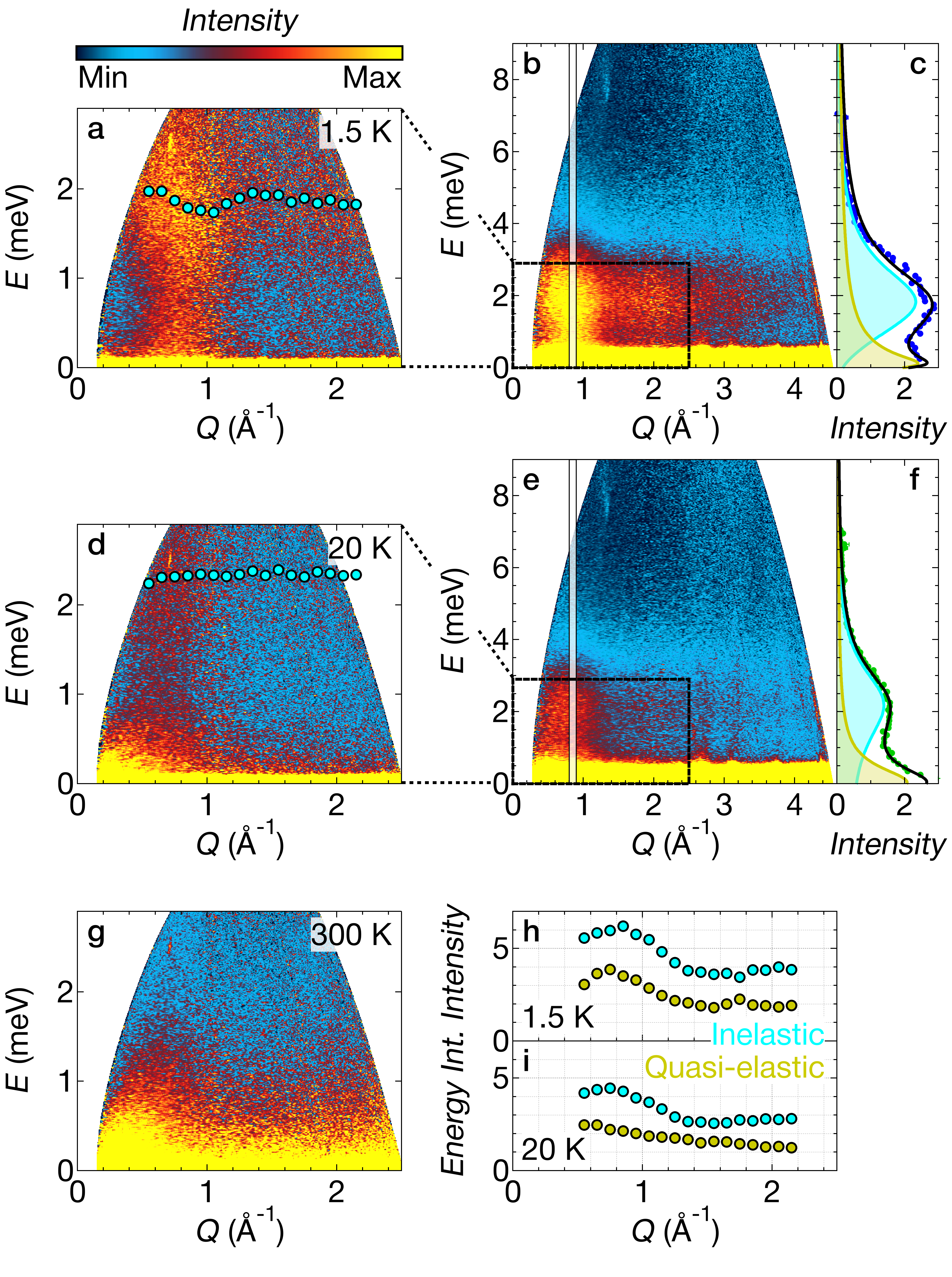 }
\caption{\textbf{a}, \textbf{d}, \textbf{g} Color maps of the neutron intensity measured with an incident neutron beam energy of $E_{\mrm{i}}=\SI{3.7}{\milli\electronvolt}$ on a powder sample at (a) $T=\SI{1.5}{\kelvin}$, (d) $T=\SI{20}{\kelvin}$, and (g) $T=\SI{300}{\kelvin}$. Light blue dots on panels (a,d) represents the characteristic energy $E_{\mrm{c}}$ of the inelastic contribution (see below). \textbf{b}, \textbf{e} Similar maps measured with $E_{\mrm{i}}=\SI{12.12}{\milli\electronvolt}$ at (b) $T=\SI{1.5}{\kelvin}$, and (e) $T=\SI{20}{\kelvin}$. The momentum and energy ranges for the measurements shown in panels (a) and (d) are indicated by the black box. \textbf{c}, \textbf{f} Constant momentum cuts integrated over the momentum range $\SIrange{0.8}{0.9}{\angstrom^{-1}}$ (white boxes in panels (b,e)) at (c) $T=\SI{1.5}{\kelvin}$ and (f) $T=\SI{20}{\kelvin}$. Cuts are fitted by a quasi-elastic (yellow) and an inelastic (light blue) contributions of characteristic energy $E_{\mrm{c}}$ (see text). \textbf{h}, \textbf{i} Momentum evolution of the energy integrated intensities of the quasi-elastic (yellow) and inelastic (light blue) contributions at (h) $T=\SI{1.5}{\kelvin}$ and (i) $T=\SI{20}{\kelvin}$, derived from the fits of panels (c) and (f) and similar ones at other momentum values.}
\label{fig:3}
\end{figure}

\subsection{Inelastic neutron scattering: dual quasi-elastic and inelastic response}
\vspace{-0.4cm}

We investigated the spin excitation spectra of a powder sample of \NCCVO\ with inelastic neutron scattering measurements recorded on LET at ISIS \cite{LEtdata} (Fig. \ref{fig:3}). Data at $T=\SI{1.5}{\kelvin}$ with an incident neutron energy of $E_{\mrm{i}} = \SI{3.7}{\milli\electronvolt}$ and $E_{\mrm{i}} = \SI{12.12}{\milli\electronvolt}$ are presented in Figs.~\hyperref[fig:3]{\ref*{fig:3}a} and ~\hyperref[fig:3]{\ref*{fig:3}b}. A very broad excitation is visible in the dynamical neutron scattering function $S(Q,E)$ below $\SI{3}{\milli\electronvolt}$, while no scattered intensity is detected above, up to $\SI{9}{\milli\electronvolt}$. The overall intensity of this large feature decreases with increasing $Q$ (indicating a magnetic origin), is gapless within our energy resolution ($\Delta E \simeq \SI{50}{\micro\electronvolt}$) near the softening wavevector $Q= \SI{0.85}{\angstrom^{-1}}$, and exhibits dispersive behavior. As the temperature increases, the intensity decreases rapidly, essentially vanishing at $T = \SI{20}{\kelvin}$ (see Figs.~\hyperref[fig:3]{\ref*{fig:3}d}, \hyperref[fig:3]{\ref*{fig:3}e} and \hyperref[fig:3]{\ref*{fig:3}f}), while the scattering becomes quasi-elastic at $T = \SI{300}{\kelvin}$ in the paramagnetic and uncorrelated regime (see Figs.~\hyperref[fig:3]{\ref*{fig:3}g} and \ref{fig:SM_Fig2} \cite{SM}). This provides clear evidence that the excitation spectrum observed at low temperatures is intrinsically linked to the magnetic correlations.

The broad excitation observed at $T = \SI{1.5}{\kelvin}$ (Fig. \ref{fig:3}) is intriguing. On one hand, there is no evidence so far for incipient magnetic order down to $T = \SI{50}{\milli\kelvin}$, which could give rise to highly damped spin waves and explain its dispersive nature. On the other hand, it does resemble the broad, sometimes dispersive, inelastic feature often seen in gapless quantum spin liquid candidates \cite{Fak2012,Chillal2020,Banerjee2016,Fujihala2020}. The existence of Na/Ca site disorder is expected to broaden any excitations, and thus care should be taken before any interpretation. However, we note that we do not observe here a completely featureless, continuum-like, spectrum. This rather suggests a moderate impact of disorder on the inelastic spectrum.

To gain a better understanding, we quantitatively characterize it by fitting successive energy cuts, integrated in a momentum window $Q\pm \SI{0.05}{\angstrom^{-1}}$ for $T=\SI{1.5}{\kelvin}$ and $T=\SI{20}{\kelvin}$ (cuts for $Q=\SI{0.85}{\angstrom^{-1}}$ are shown in Figs.~\hyperref[fig:3]{\ref*{fig:3}c},~\hyperref[fig:3]{\ref*{fig:3}f}, \hyperref[fig:SM_Fig4]{\ref*{fig:SM_Fig4}a} and \hyperref[fig:SM_Fig4]{\ref*{fig:SM_Fig4}b} \cite{SM}). Clearly, the data cannot be described by a single peak of any shape due to the dip in intensity around $E \approx \SI{1}{\milli\electronvolt}$. Instead, a dual response is necessary to accurately reproduce the signal. We achieve very good agreement with the data up to at least $Q =\SI{2.25}{\angstrom^{-1}}$ using the scattering function:
\begin{equation}
\begin{gathered}
S(E,T) = \frac{1}{1-\exp(-E/k_{\mrm{B}}T)}\left[\chi''_{\mrm{Qel}}(E)+\chi''_{\mrm{Inel}}(E)\right];\\
\begin{aligned}
\chi''_{\mrm{Qel}}(E) &= \frac{z\gamma E}{E^{2}+\gamma^{2}};\\
\chi''_{\mrm{Inel}}(E) &=  \left[\frac{Z\Gamma}{(E-E_{\mrm{c}})^{2}+\Gamma^{2}}-\frac{Z\Gamma}{(E+E_{\mrm{c}})^{2}+\Gamma^{2}}\right].
\end{aligned}
\label{eq:Escans}
\end{gathered}
\end{equation}
The first term of Eq. \ref{eq:Escans} represents a quasi-elastic signal of amplitude $z$. Its linewidth $\gamma$ (Fig.~\hyperref[fig:SM_Fig4]{\ref*{fig:SM_Fig4}c} \cite{SM}) is nearly constant with $Q$ but double with temperature: $\gamma\sim\SI{0.25}{\milli\electronvolt}$ at $T=\SI{1.5}{\kelvin}$ and $\gamma\sim\SI{0.49}{\milli\electronvolt}$ at $T=\SI{20}{\kelvin}$. This accounts for the gapless part of the scattering, with a characteristic frequency $\nu = \gamma / h \simeq \SI{60}{\giga\hertz}$, consistent with the fast fluctuations observed in $\mu$SR. The second term corresponds to an inelastic contribution of amplitude $Z$, characteristic energy $E_{\mrm{c}}$ and linewidth $\Gamma$. As revealed by the momentum dependence of $E_{\mrm{c}}$ (shown in light blue dots in Figs.~\hyperref[fig:3]{\ref*{fig:3}a} and \hyperref[fig:3]{\ref*{fig:3}d}), this signal is dispersive  at $T=\SI{1.5}{\kelvin}$ and becomes dispersionless already at $T=\SI{20}{\kelvin}$, reflecting weaker magnetic correlations. On the contrary, the linewidth $\Gamma$ (Fig.~\hyperref[fig:SM_Fig4]{\ref*{fig:SM_Fig4}c} \cite{SM}) remains nearly constant over both $Q$ and the temperature $T$: $\Gamma\sim\SI{1.2}{\milli\electronvolt}$ at $T=\SI{1.5}{\kelvin}$ and $\Gamma\sim\SI{1.4}{\milli\electronvolt}$ at $T=\SI{20}{\kelvin}$. In Figs.~\hyperref[fig:3]{\ref*{fig:3}a} and \hyperref[fig:3]{\ref*{fig:3}d}, we show the momentum modulation of the energy integrated intensity of the two excitations (defined as the energy integration of each contributions to $S(E,T)$ in the range $\SIrange{0}{9}{\milli\electronvolt}$, see SM Sec. E \cite{SM}). At $\SI{1.5}{\kelvin}$, both the quasi-elastic and inelastic energy integrated intensities closely track the intensity modulation observed in polarized neutron diffraction (Fig.~\hyperref[fig:1]{\ref*{fig:1}b}). Notably, the maximum integrated intensity occurs at $Q \sim \SI{0.9}{\angstrom^{-1}}$, with a second local maximum observed for $Q \sim \SI{2}{\angstrom^{-1}}$. Finally, the momentum-dependent modulation observed in Figs.~\hyperref[fig:3]{\ref*{fig:3}h} and \hyperref[fig:3]{\ref*{fig:3}i} clearly indicates a dispersive behavior. However, we were not able to resolve any specific inelastic mode within the energy resolution of our powder spectra. 

This time-of-flight data with energy resolution shows that a significant portion of the energy-integrated signal observed on the D7 spectrometer is indeed quasi-elastic and inelastic, suggesting that spin dynamics still persist on a much shorter timescale than that probed by $\mu$SR. The dual response in inelastic neutron scattering thus naturally stems from: (1) fast fluctuating $S=1/2$ moments leading to a quasi-elastic contribution, consistent with the spin dynamics probed by the muon, and (2) short-live dispersive excitations extending up to $\sim 4$~meV, either linked to highly-damped spin-waves reminiscent of a nearby order or to unconventional spinon excitations of a QSL phase \cite{Ghosh2019,Sonnenschein2020}. 



\subsection{High-temperature series expansion: strongly frustrated spin chains}
\vspace{-0.4cm}

\begin{figure}
\includegraphics[width=1\columnwidth]{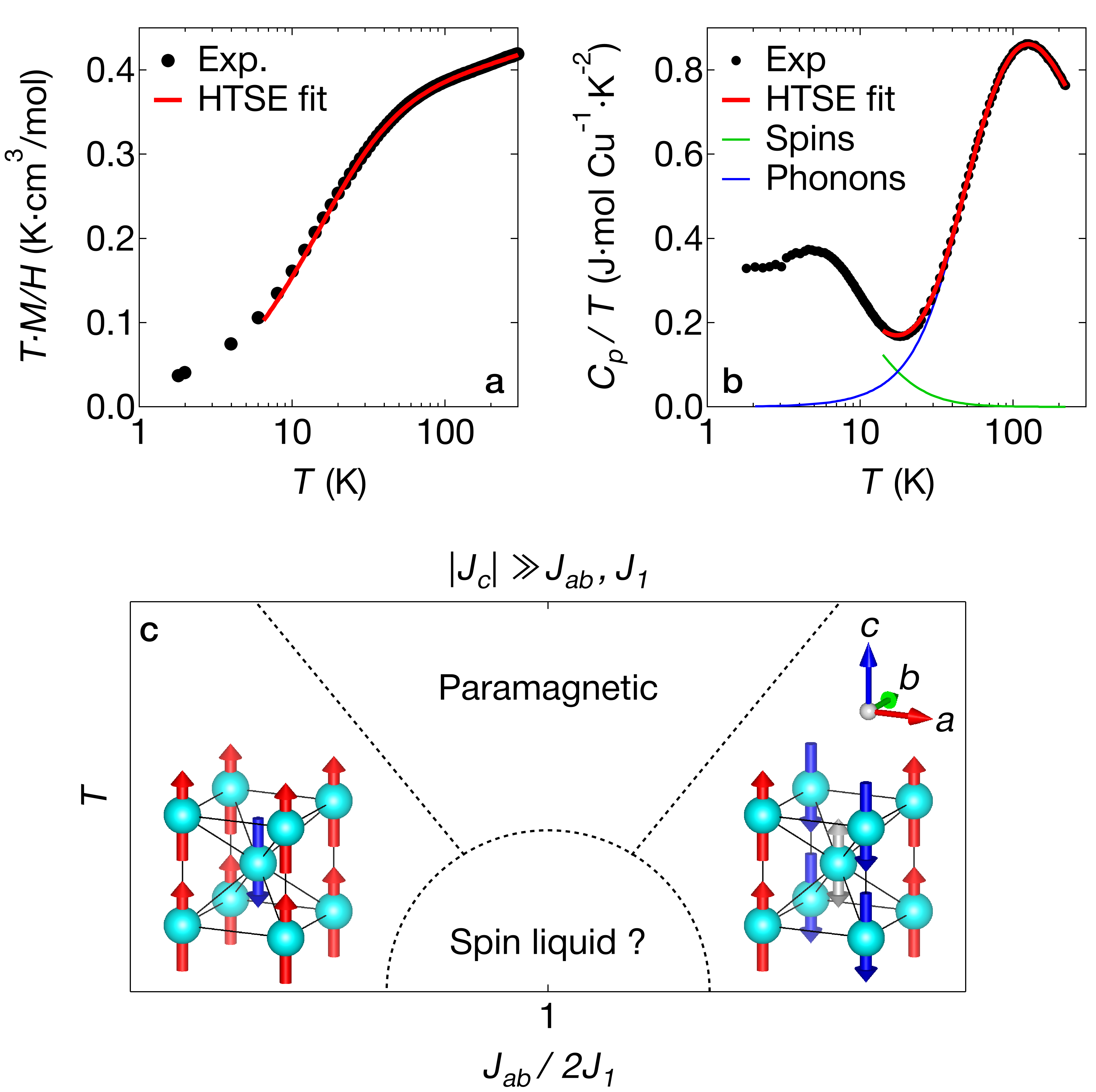 }
\caption{\textbf{a} Comparison of the magnetic susceptibility measured with $\mu_{0}H=\SI{5}{\tesla}$ (black dots) and the HTSE fit down to $T=\SI{10}{\kelvin}$ (red line). \textbf{b} Total specific heat measured in zero-field (black dots), with the model (red line) consisting of a spin contribution (green line) and a phononic one (blue line) fitted from the HTSE down to $T=\SI{10}{\kelvin}$. \textbf{c} Schematic phase diagram of the spin model of Eq. \ref{eq:SpinModel} for a dominant ferromagnetic $J_{c}$. The exchange parameters of \NCCVO\ extracted from the HTSE place it near $J_{ab}/2J_{1}\sim 1$.}
\label{fig:4}
\end{figure}

We analyse the bulk magnetic susceptibility and specific heat data with a multi-exchange Heisenberg model, using high temperature series expansions (HTSE). The $12^{\mrm{th}}$-order cubic $J_1-J_2$ HTSE (first and second nearest-neighbors) failed to reproduce the data, not improved by including the third nearest-neighbors in a $J_1-J_2-J_3$ model. A better agreement is achieved by taking into account the tetragonal distorsion along the $c$ axis and including couplings up to the second nearest-neighbors (now separated between those along $a,b$ and those along $c$), namely
\begin{equation}
\hat{\mathcal{H}} = J_{1}\sum_{\left<i,j\right>}\hat{\bf{S}}_{i}\cdot\hat{\bf{S}}_{j} + J_{c}\sum_{\left<\left<i,j\right>\right>_{c}}\hat{\bf{S}}_{i}\cdot\hat{\bf{S}}_{j} + J_{ab}\sum_{\left<\left<i,j\right>\right>_{a,b}}\hat{\bf{S}}_{i}\cdot\hat{\bf{S}}_{j}.
\label{eq:SpinModel}
\end{equation}
In Eq. \ref{eq:SpinModel}, $\left<i,j\right>$ denote first nearest-neighbors ($1^{\mrm{st}}$ NN), while $\left<\left<i,j\right>\right>_{a,b,c}$ corresponds to second nearest-neighbor ($2^{\mrm{nd}}$ NN) along the $a$, $b$ or $c$ axis. The data were reproduced down to $T=\SI{10}{\kelvin}$ using Pade approximant of the $12^{\mrm{th}}$-order HTSE for the susceptibility and $13^{\mrm{th}}$-order HTSE for the specific heat (Figs.~\hyperref[fig:4]{\ref*{fig:4}a} and \hyperref[fig:4]{\ref*{fig:4}b}). Based on our best fit, the exchange parameters are $J_{1}=\SI{4.5 \pm 0.5}{\kelvin}$, $J_{ab}=\SI{7.5\pm 0.5}{\kelvin}$, and $J_{c}=\SI{-22\pm 2}{\kelvin}$ (See SM Sec. F \cite{SM} for details on the procedure and uncertainties). Clearly, the small tetragonal distortion greatly affects the magnetic properties of \NCCVO: the strongly ferromagnetic $J_{c}$ interaction promotes ferromagnetic spin chains along the $c$ axis, yet frustrated by the competing antiferromagnetic $J_{1}$ and $J_{ab}$ couplings.

This result is consistent with our polarized neutron diffraction measurements. Indeed, we found AF correlations for $1^{\mrm{st}}$ NN and weaker F correlations for $2^{\mrm{nd}}$ NN, reflecting the partial averaging of strong F exchange along the $c$ axis and weaker AF exchange along the $a,b$ axis.

\section{Discussion}

The $J_1 - J_2$ $S=1/2$ Heisenberg bcc model is known to host a quantum phase transition between two antiferromagnetic orders when the first NN interaction $J_1$ and the second NN neighbour interaction $J_2$ compete for $J_2 / J_1 \simeq 0.7$~\cite{Oitmaa2004}. So far, theoretical works have concluded to the absence of quantum spin liquid phase in the vicinity of such critical point \cite{Oitmaa2004, Farnell2016,Majumdar_2009}. However, adding a third NN interaction $J_3$ can stabilize an extended quantum disordered phase according to a recent pseudofermion functional renormalization group study \cite{Ghosh2019}.

Here, our minimal model Hamiltonian includes three exchange interactions $J_{1} - J_{ab} - J_{c}$ (Eq. \ref{eq:SpinModel}). Our exchange values suggest significant magnetic frustration between $J_{1}$ and $J_{ab}$. Indeed, assuming ferromagnetic chains along the c axis, two magnetic orders (illustrated in Fig.~\hyperref[fig:4]{\ref*{fig:4}c}) compete. If the coupling $J_{1}$ between the two sublattices dominates, both cubic sublattices order ferromagnetically while the order between them is antiferromagnetic. Conversely, if the interchain coupling $J_{ab}$ is stronger, neighboring chains within each sublattice order antiferromagnetically while the mean field of one sublattice on the other one cancels out. Crucially, because each spin is surrounded by $8$ others coupled with $J_{1}$ and $4$ others coupled with $J_{ab}$, the ratio $J_{ab}/2J_{1}$ determines the balance of the two interactions. In our case, this value is rather close to unity ($J_{ab}/2J_{1} \simeq 0.76$), highlighting the strong frustration among these competing  exchange paths. Suprinsingly, while the model Hamiltonian relevant for \NCCVO\ significantly departs from the $J_1 - J_2$ Heisenberg bcc model due to the tetragonal distortion, we recover an analogous strongly frustrated situation. Although a direct comparison with the bcc calculations of Refs. \cite{Ghosh2019,Sonnenschein2020} is difficult due to the distorsion present in \NCCVO, we note that the quantum paramagnetic region of their phase diagram lies between the two magnetic orders described above (corresponding to those identified as $(2\pi,0,0)$ and $(\pi,\pi,0)$). Thus, a potentially frustration-induced spin liquid state of \NCCVO\, as suggested in the schematic phase diagram in Fig.~\hyperref[fig:4]{\ref*{fig:4}c}, might be related to their quantum disordered state. Along this line, a comparison to the excitation spectrum of quantum spin liquid phases potentially existing for this $J_{1} - J_{ab} - J_{c}$ interaction scheme could shed new lights on the unconventional inelastic and quasi-elastic signals observed in \NCCVO. Finally, although the intrinsic disorder from mixed Na/Ca site occupancy appears less severe than in some rare-earth quantum magnets \cite{Zhu2017}, it should still be accounted for in any future realistic simulations of the excitations. In particular, the link between persistent fluctuations as observed in $\mu$SR and the presence of disorder, even at a minute level \cite{Hodges2002,DOrtenzio2013,Chang2014}, should be investigated further.

The presence of strong 1D correlations in \NCCVO\ was already pointed out in early studies \cite{Kazei1983} based on the analysis of bulk macroscopic measurements. However, these correlations were believed to be antiferromagnetic in nature. This interpretation was difficult to reconcile with simple theoretical models, and the qualitative disagreement between the data and the models was attributed to correlated chains of variable lengths due to complex Jahn-Teller effects. Our analysis provides new insight into this issue, offering compelling evidence for ferromagnetically correlated chains along the $c$ axis, which are coupled antiferromagnetically to the neighboring chains of both sublattices. However, as demonstrated by our $\mu$SR measurements, spin-lattice effects are at play down to the lowest measured temperature of $T = \SI{50}{\milli\kelvin}$. The orbital overlap - and hence the exchange coupling - could still vary well below $T=\SI{10}{\kelvin}$, highlighting the need for comprehensive studies of the strongly magnetically correlated state of \NCCVO\ at dilution temperatures and DFT calculations to confirm the frustrated exchange scheme. An exciting challenge would be to further control the lattice distortions through applied pressure or ultra-fast terahertz pulses, as recently realized in SrCu$_2$(BO$_3$)$_2$ \cite{Giorgianni2023}, and tune the spin-lattice interaction responsible for the spin liquid behavior in \NCCVO.

In summary, we demonstrated using a combination of macroscopic measurements, neutron scattering and muon spin relaxation techniques, that the nearly cubic, magnetically correlated, $S=1/2$ system \NCCVO\ shows persistent spin dynamics down to at least $T=\SI{50}{\milli\kelvin}$ while it becomes correlated in a temperature range two orders of magnitude higher. The magnetic interactions are determined from high-temperature series expansion of the magnetic susceptibility and the specific heat, yielding three relevant exchange interactions $J_{1}=\SI{4.5 \pm 0.5}{\kelvin}$, $J_{ab}=\SI{7.5\pm 0.5}{\kelvin}$, and $J_{c}=\SI{-22\pm 2}{\kelvin}$, responsible for correlated ferromagnetic chains as well as strong frustration between them. The existence of a spin-liquid phase for such parameters would need to be explored theoretically. 
Alternatively, the interplay between Jahn-Teller vibrations and magnetic frustration via spin-lattice coupling could also provide an appealing explanation for the absence of magnetic order down to $\SI{50}{\milli\kelvin}$ and the stabilization of the intriguing three-dimensional spin liquid behavior of \NCCVO.


\section{Acknowledgments}
Y. A. thanks V. Simonet and E. Lhotel for fruitful discussions. E. K. acknowledges very useful discussions with H. O. Jeschke and Y. Iqbal during the HFM2024 conference. Experiments at the ISIS Neutron and Muon Source were supported by beamtime allocations RB2410480 and RB2210305 from the Science and Technology Facilities Council. We acknowledge technical support during our experiment performed at the ILL (proposal 5-32-927). This work was supported by the French Agence Nationale de la Recherche, under Grant No. ANR- 18-CE30-0022 “LINK”. E. K. and R. S. acknowledge financial support from the labex PALM for the QuantumPyroMan project (ANR-10-LABX-0039-PALM).

%


\clearpage

\renewcommand{\thefigure}{S\arabic{figure}} 
\renewcommand{\theequation}{S\arabic{equation}} 
\renewcommand{\thetable}{S\arabic{table}}
\renewcommand*{\thesubsection}{\Alph{subsection}}

\setcounter{figure}{0} 
\setcounter{equation}{0} 
\setcounter{table}{0}
\setcounter{section}{0}

\makeatletter
\renewcommand{\theHfigure}{supp.\arabic{figure}}
\makeatother

\title{Supplemental Material for "Evidence for spin liquid behavior in the frustrated three-dimensional  $\mathbf{S = 1/2}$ Heisenberg garnet NaCa$_{\mathbf{2}}$Cu$_{\mathbf{2}}$(VO$_{\mathbf{4}}$)$_{\mathbf{3}}$"}

\maketitle

\onecolumngrid

\section{A. Sample synthesis and cristallographic structure}
\label{SMA}

In order to prepare the polycrystalline samples of \NCCVO, high purity oxide materials of Na$_{2}$CO$_{3}$, CaCO$_{3}$, CuO, and V$_{2}$O$_{5}$ were mixed in stoichiometry and the mixture was pelletized and heat treated in the temperature range $\SIrange{500}{720}{\celsius}$ for twelve hours at each temperature with several intermittent grindings. 

\begin{figure*}[h!]
\includegraphics[width=\textwidth]{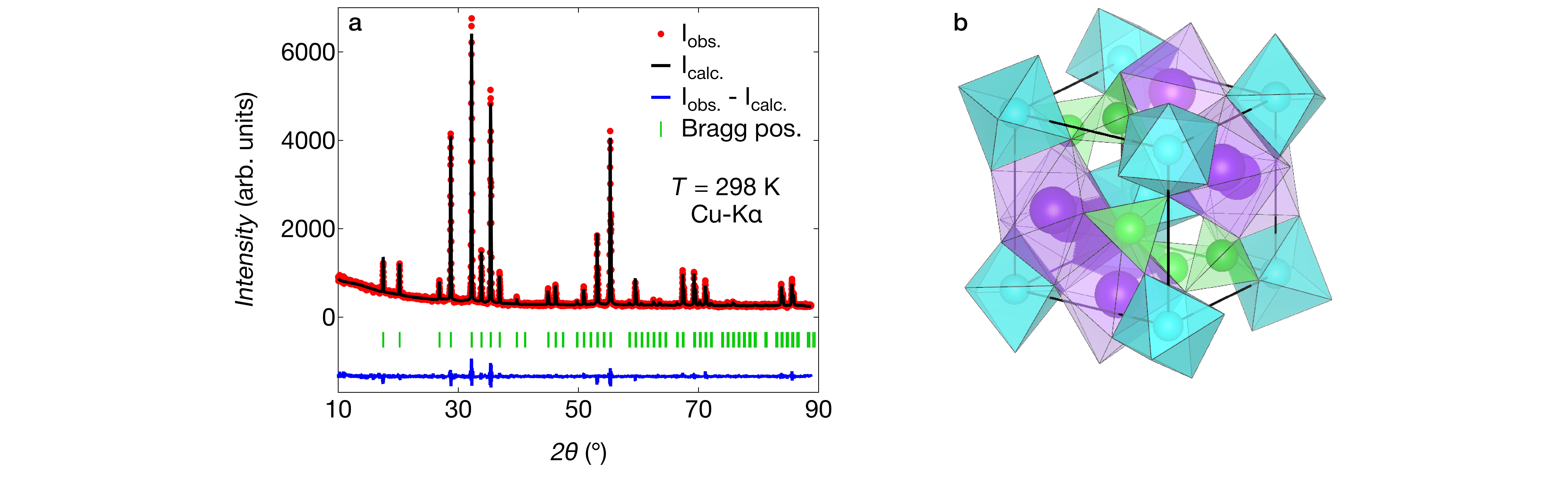 }
\caption{\textbf{a} X-ray diffractograms of polycrystalline \NCCVO\ at room temperature (red dots). Rietveld refinement of the data in the space group n$^{\circ}$230 Ia$\bar{3}$d is shown in black and the difference between refinement and data in blue. The Bragg peak positions are indicated by green ticks. \textbf{b} $1/8$ of the cubic unit cell of \NCCVO. Na$^{+}$/Ca$^{2+}$ (24c), Cu$^{2+}$ (16a) and V$^{5+}$ (24d) ions are pruple, blue, and green respectively. Oxygens O$^{2-}$ (96h) landing at the vertices of the different polyhedra are not shown explicitely.}
\label{fig:SM_Fig1}
\end{figure*}

The x-ray diffractogram of our \NCCVO\ polycrystalline sample at $T=\SI{298}{\kelvin}$ is shown in Fig.~\hyperref[fig:SM_Fig1]{\ref*{fig:SM_Fig1}a}. The refinement (Bragg-R factor $=4.94$) confirms the cubic garnet structure shown in Fig.~\hyperref[fig:SM_Fig1]{\ref*{fig:SM_Fig1}b} (space group n$^{\circ}$230 Ia$\bar{3}$d, $8$ formula per unit cell). We found a unit cell parameter $a=b=c=\SI{12.425}{\angstrom}$, in very good agreement with previous studies \cite{Bayer1965,Kazei1982,Kazei1983}. Other refined free parameters and description of the structure are given in Table \ref{tab:structure}. 

\begin{table}[h!]
\begin{ruledtabular}
\begin{tabular}{c|c|c|c|c}
Atom & Na / Ca & Cu & V & O \\
\hline\noalign{\vskip 0.5mm} 
Wyckoff & $24c$ & $16a$ & $24d$ & $96h$ \\
\hline\noalign{\vskip 0.5mm} 
Symm. & $222$ & $\bar{3}$ & $\bar{4}$ & $1$ \\
\hline\noalign{\vskip 0.5mm} 
$x$ & $0$ & $0$ & $0$ & $-0.0394$ \\
$y$ & $1/4$ & $0$ & $1/4$ & $0.0541$ \\
$z$ & $1/8$ & $0$ & $3/8$ & $0.1563$\\
\hline\noalign{\vskip 0.5mm} 
Occ. & $0.333$ / $0.667$ & $1$ & $1$ & $1$\\
\hline\noalign{\vskip 0.5mm}
Poly. &  Dodecahedron & Octahedron & Tetrahedron & n.a.
\end{tabular}
\label{TableStruct}
\end{ruledtabular}
\caption{Results of the Rietvelt refinement of \NCCVO\ X-ray diffraction data in the Ia$\bar{3}$d (n$^{\circ}$230) space group.}
\label{tab:structure}
\end{table}  

\FloatBarrier

\section{B. Complement on polarized neutron diffraction}
\label{SMB}

\begin{figure*}[h!]
\includegraphics[width=\textwidth]{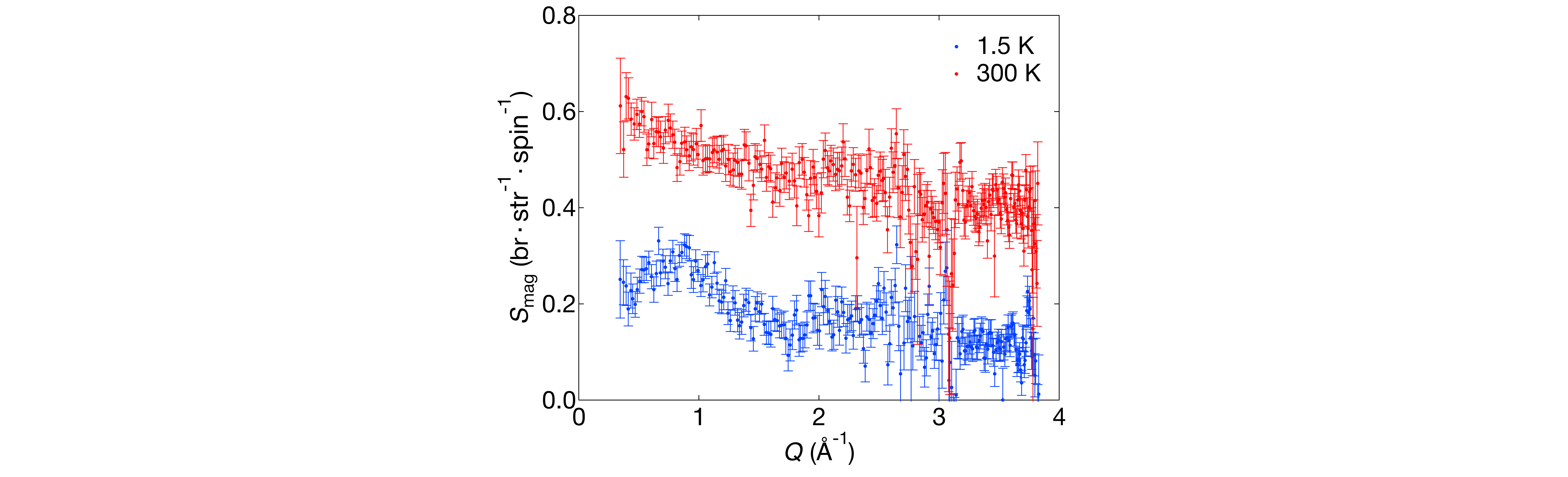 }
\caption{Magnetic scattering intensity measured with polarized neutron diffraction. Data at $T=\SI{300}{\kelvin}$ are offset by $0.3$.}
\label{fig:SM_Fig2}
\end{figure*}

We recorded polarized neutron scattering data on the D7 spectrometer at the ILL \cite{D7}. A powder sample of $\SI{3.598}{\gram}$ was loaded into an aluminium annular can with a $\SI{20}{\milli\meter}$ outer diameter and $\SI{19}{\milli\meter}$ inner diameter. We operate in the diffraction mode, using an incident neutron wavelength of $\SI{3.1}{\angstrom}$ ($E_{\mrm{i}}=\SI{8.5}{\milli\electronvolt}$) and NSF:SF counting ratio of 1:4 for the three X, Y and Z polarization directions. In Fig.~\ref{fig:SM_Fig2}, we present a comparison of the measurements at $T=\SI{1.5}{\kelvin}$ and $T=\SI{300}{\kelvin}$. Low temperatures data in Fig.~\hyperref[fig:1]{\ref*{fig:1}b} corresponds to the 2-points average of these ones. Oscillations in the intensity are cleary visible at low temperatures, but are no longer present at room temperature. This indicates that the system is uncorrelated (paramagnetic) at $T=\SI{300}{\kelvin}$.

\begin{figure*}[h!]
\includegraphics[width=\textwidth]{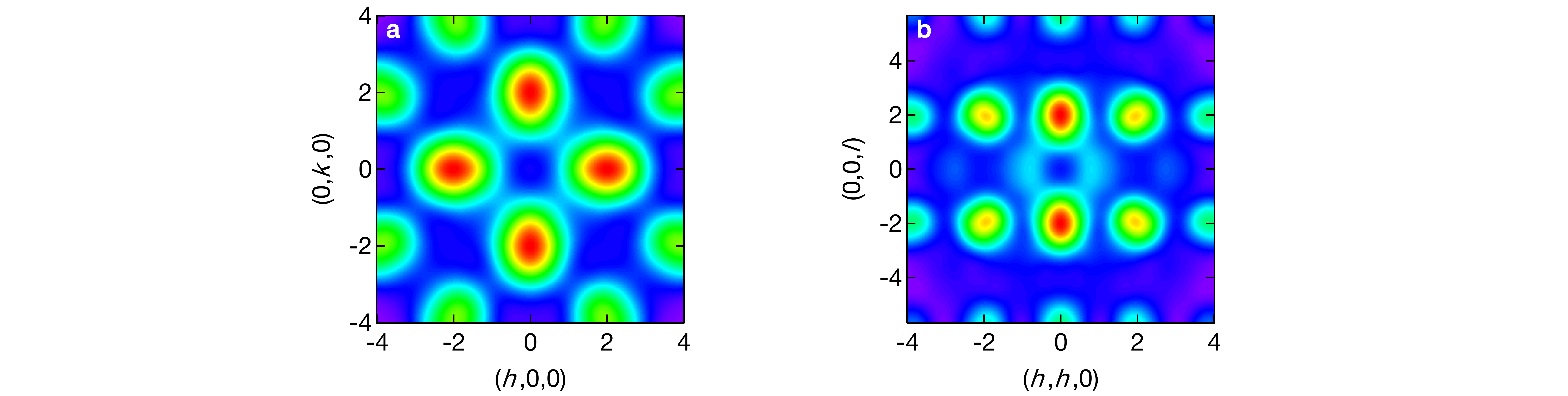 }
\caption{Expected single crystal magnetic scattering calculated with \textsc{SCATTY} in \textbf{a} the $(hk0)$ plane and \textbf{b} the $(hhl)$ plane.}
\label{fig:SM_Fig3}
\end{figure*}

The expected single crystal magnetic diffuse scattering of \NCCVO\ is shown in Figs.~\hyperref[fig:SM_Fig3]{\ref*{fig:SM_Fig3}a} and \hyperref[fig:SM_Fig3]{\ref*{fig:SM_Fig3}b} in the $(hk0)$ and $(hhl)$ scattering planes, respectively. These patterns were calculated using the software \textsc{SCATTY}, based on the Monte-Carlo simulations of the powder data obtained with \textsc{SPINVERT}.

\section{C. Specific heat data analysis}
\label{SMC}

We determined the bcc lattice contribution to the specific heat of \NCCVO\ from measurements on the non-magnetic analogue \NCMVO\ (see Ref. \cite{Hardy2003} for the method). To account for the tetragonal distortion induced by Jahn-Teller effects below $\SI{250}{\kelvin}$, and to isolate the signal of magnetic origin, we subtracted the bcc lattice contribution from the measured specific heat divided by temperature. The resulting signal was fitted using two phenomenological log-normal distributions (see Fig.~\hyperref[fig:1]{\ref*{fig:1}g}), representing the magnetic and Jahn-Teller (non-bcc lattice) parts, respectively. Each distribution was defined as
\begin{equation}
f(T) = \frac{A}{T\sigma\sqrt{2\pi}}\exp\left(\frac{-(\ln(T)-\mu)^{2}}{2\sigma^{2}}\right)
\end{equation}
where $A$ is the area and $\sigma$, $\mu$ two fit parameters. These distributions were choosen for their asymmetry and vanishing limits in both low and high temperatures. For the magnetic part, our fits gives $A_{\mrm{Mag}} = 5.50$, $\sigma_{\mrm{Mag}} = 0.92$ and $\mu_{\mrm{Mag}} = 2.25$. For the Jahn-Teller part, we obtained $A_{\mrm{JT}} = 3.05$, $\sigma_{\mrm{JT}} = 0.47$ and $\mu_{\mrm{JT}} = 4.26$.

\begin{figure*}[ht!]
\includegraphics[width=\textwidth]{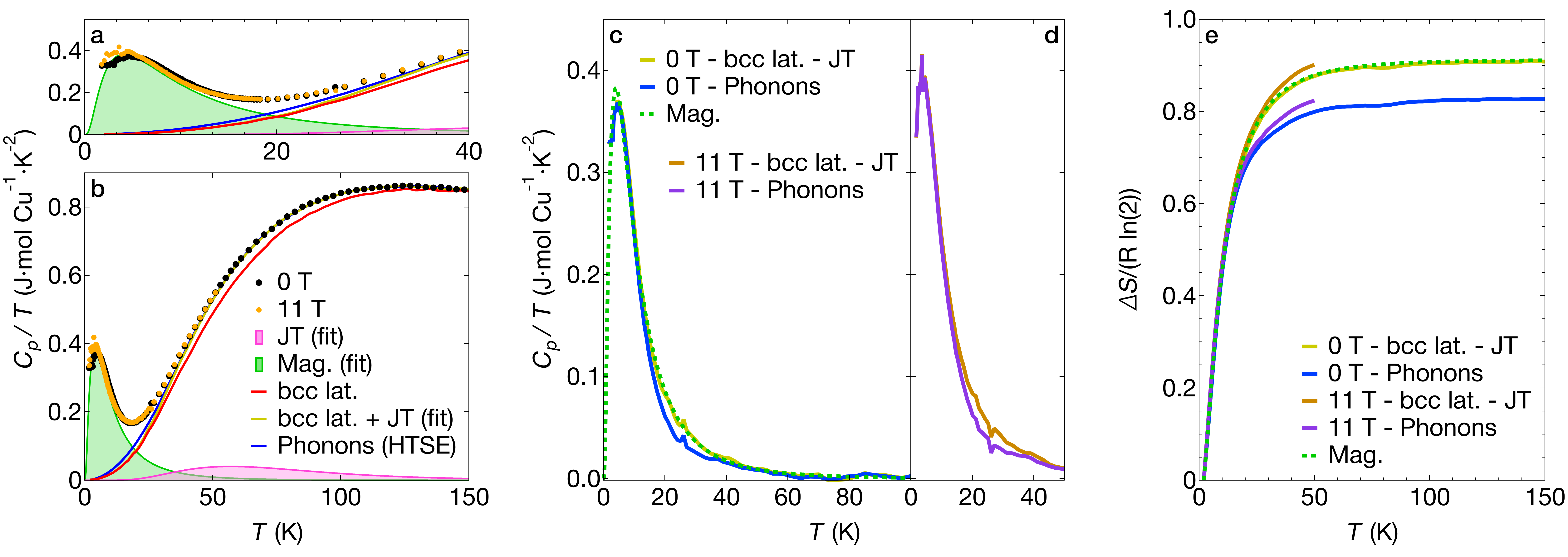 }
\caption{\textbf{a}, \textbf{b} Temperature dependence of the specific heat of \NCCVO\ at $\SI{0}{\tesla}$ and $\SI{11}{\tesla}$ (dots), fitted Jahn-Teller and magnetic parts (shaded areas), and estimations of the non-magnetic contributions to the specific heat (colored lines). \textbf{c}, \textbf{d} Magnetic specific heat obtained using the different estimations of the non-magnetic contribution. \textbf{e} Integrated entropy from $\SI{2}{\kelvin}$.}
\label{fig:SM_Fig6}
\end{figure*}

Figs.~\hyperref[fig:SM_Fig6]{\ref*{fig:SM_Fig6}a} and \hyperref[fig:SM_Fig6]{\ref*{fig:SM_Fig6}b} compare the total specific heat of \NCCVO\ (measured at \SI{0}{\tesla} and \SI{11}{\tesla}) with the bcc lattice contribution, the two fitted components, and the two approximations for the total non-magnetic specific heat used in the main text: (i) the sum of the bcc lattice and fitted Jahn-Teller parts (Fig.~\ref{fig:1}), and (ii) the calculated phonon contribution (Fig.~\ref{fig:4}, see part. F below). These two approximations agrees above $\SI{40}{\kelvin}$ but slightly deviates at lower temperatures. A comparison of the magnetic specific heat obtained by subtracting our measurements with the two different non-magnetic signals at $\SI{0}{\tesla}$ and $\SI{11}{\tesla}$ are shown in Figs.~\hyperref[fig:SM_Fig6]{\ref*{fig:SM_Fig6}c} and \hyperref[fig:SM_Fig6]{\ref*{fig:SM_Fig6}d}, respectively. The integrated entropy from $\SI{2}{\kelvin}$ (Fig.~\hyperref[fig:SM_Fig6]{\ref*{fig:SM_Fig6}e}) reaches $\SI{83}{\percent}$ and $\SI{90}{\percent}$ of $R\ln(2)$ for $\mu_{0}H=\SI{0}{\tesla}$ and slightly more for $\mu_{0}H=\SI{11}{\tesla}$. The $\SI{5}{\percent}$ difference from the value reported in the main text arises from the lower integration limit, down to $\SI{0}{\kelvin}$.

\section{D. Muon spin relaxation}
\label{SMD}

The muon spin relaxation measurements were carried out at the ISIS, Neutron and Muon facility, UK, on the MuSR spectrometer. The powder sample ($\SI{1}{\gram}$) was mixed with a small amount of GE-varnish and placed on an $\SI{3}{\centi\meter}$ silver plate and loaded into a dilution fridge. A $\SI{25}{\micro\meter}$ silver-foil degrader was added before the muon beam. 


\section{E. Inelastic neutron scattering and fits of the constant momentum cuts}
\label{SME}

The inelastic neutron scattering measurements were performed on the LET time-of-flight  spectrometer at the ISIS, STFC, Neutron and Muon source. A powder sample of $\SI{3.503}{\gram}$ was loaded into an aluminium annular can with a $\SI{16}{\milli\meter}$ outer diameter and $\SI{14}{\milli\meter}$ inner diameter. We operate using the multiplexing chopper configuration allowing the simultaneous measurements of three different incident energies within a single time frame: $E_\mrm{i} = 12.12$, $3.7$, $\SI{1.77}{\milli\electronvolt}$. Typical measurement for a given temperature last for $\SI{4}{\hour}$.

\begin{figure*}
\includegraphics[width=1\textwidth]{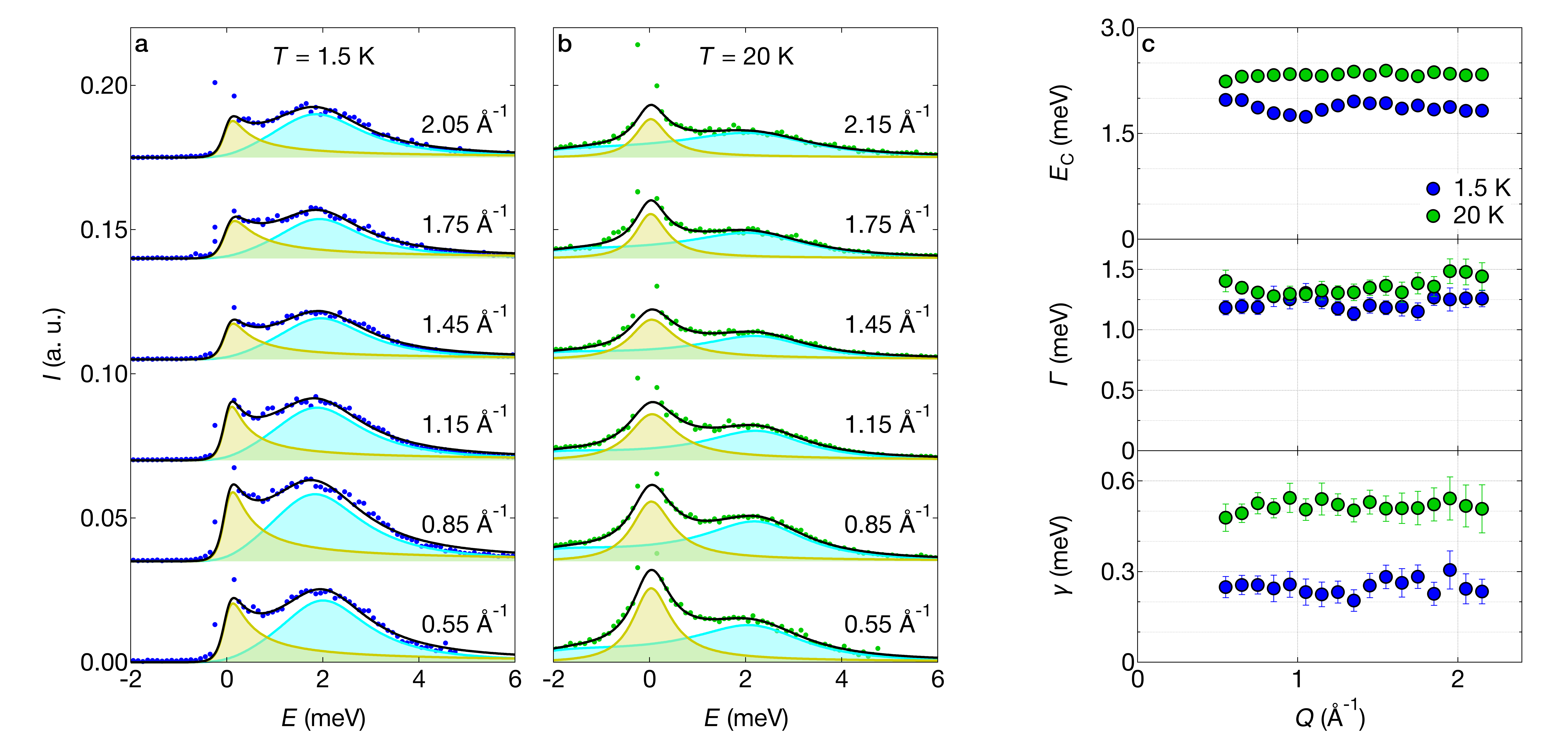 }
\caption{\textbf{a}, \textbf{b} Constant momentum cuts obtained from inelastic neutron scattering measurements (dots) with their fits (black line), consisting of a quasi-elastic contribution (yellow area) and an inelastic contribution (blue area) at (a) $T=\SI{1.5}{\kelvin}$ and (b) $T=\SI{20}{\kelvin}$. \textbf{c} Fit parameters: characteristic energy of the inelastic contribution (top), half width at half maximum of the inelastic contribution (middle) and half width at half maximum of the quasi-elastic contribution (bottom).}
\label{fig:SM_Fig4}
\end{figure*}

The constant momentum cuts (shown in Figs.~\hyperref[fig:3]{\ref*{fig:3}c}, ~\hyperref[fig:3]{\ref*{fig:3}f} and \ref{fig:SM_Fig4}) were obtained by combining the data measured with  $E_{\mrm{i}} = \SI{3.7}{\milli\electronvolt}$ and  $E_{\mrm{i}} = \SI{12.12}{\milli\electronvolt}$. We first integrated the measured intensity for both incident energies in successive windows defined by $Q\pm \SI{0.05}{\angstrom}$, with $Q$ ranging from $Q=\SI{0.55}{\angstrom^{-1}}$ to $Q=\SI{2.15}{\angstrom^{-1}}$. The intensity of the cuts obtained with $E_{\mrm{i}} = \SI{3.7}{\milli\electronvolt}$ multiplied by a factor $1.8$ overlap the intensity of the cuts obtained with $E_{\mrm{i}} = \SI{12.12}{\milli\electronvolt}$. Thus, we merged the two sets of data: the resulting by combining the data from $E_{\mrm{i}} = \SI{3.7}{\milli\electronvolt}$ (multiplied by $1.8$) between $\SI{-1.6}{\milli\electronvolt}$ and $\SI{1.8}{\milli\electronvolt}$, and the data from  $E_{\mrm{i}} = \SI{12.12}{\milli\electronvolt}$ elsewhere.

As described in main text, the cuts for different $Q$ values were fitted using the scattering function
\begin{equation}
\begin{aligned}
S(E,T) &= \frac{1}{1-\exp(-E/k_{\mrm{B}}T)}\left[\chi''_{\mrm{Qel}}(E)+\chi''_{\mrm{Inel}}(E)\right]\\
&= \frac{1}{1-\exp(-E/k_{\mrm{B}}T)}\left[\frac{z\gamma_{Q}E}{E^{2}+\gamma^{2}} + \left(\frac{Z\Gamma}{(E-E_{\mrm{c}})^{2}+\Gamma^{2}}-\frac{Z\Gamma}{(E+E_{\mrm{c}})^{2}+\Gamma^{2}}\right)\right]
\label{eq:Escans_SM}
\end{aligned}
\end{equation}
which represents the sum of a quasi-elastic and an inelastic contributions. Only the positive part up to $E=\SI{6}{\milli\electronvolt}$ of the merged cuts were fitted, excluding the region $0\le E\le\SI{0.25}{\milli\electronvolt}$ dominated by the elastic peak. The fits show excellent agreement with the data at both temperatures and for all momentum values $Q$ ranging from $Q = \SI{0.55}{\angstrom^{-1}}$ to $Q = \SI{2.15}{\angstrom^{-1}}$. The parameters $E_{\mrm{c}}$, $\Gamma$ and $\gamma$ are displayed in Fig.~\hyperref[fig:SM_Fig4]{\ref*{fig:SM_Fig4}c}.


The inelastic and quasi-elastic energy integrated intensity, depicted in the main text (Figs.~\hyperref[fig:3]{\ref*{fig:3}h} and \hyperref[fig:3]{\ref*{fig:3}i}), were calculated by integrating the fitted quasi-elastic and inelastic contributions of the scattering function $S(E,T)$ between $\SI{0}{\milli\electronvolt}$ and $\SI{9}{{\milli\electronvolt}}$. Formally, for each $Q$ corresponding to a constant momentum cut, the inelastic energy integrated intensity $A_{\mrm{Inel}}$ and the quasi-elastic energy integrated intensity $A_{\mrm{Qel}}$ were defined as
\begin{equation}
\begin{gathered}
A_{\mrm{Qel}} = \int_{0}^{9} \frac{1}{1-\exp(-E/k_{\mrm{B}}T)}\chi''_{\mrm{Qel}}(E)\mrm{d}E;\\
A_{\mrm{Inel}} = \int_{0}^{9} \frac{1}{1-\exp(-E/k_{\mrm{B}}T)}\chi''_{\mrm{Inel}}(E)\mrm{d}E;\\
\end{gathered}
\end{equation}
where $\chi''_{\mrm{Qel}}(E)$ (resp. $\chi''_{\mrm{Qel}}(E)$) depends on the parameters $z$ and $\gamma_{Q}$ (resp. $Z$, $E_{\mrm{c}}$ and $\Gamma$) obtained from the fits of the constant momentum scans.\\

\section{F. High temperature series expansion}
\label{SMF}

\indent The measured linear susceptibility is defined as $\chi_l^{\rm expe}=M/\mu_{0}H$. For each set of exchange energy $\{J_\alpha\}$, the HTSE of $T\chi_l^{\rm theo}$ is evaluated and the linear parameters $A_\chi$ and $B_\chi$ are calculated by minimizing the error $E_\chi$:
\begin{equation}
	E_\chi=\sum_{T_i>T_{\min}} \left( T_i \chi_l^{\rm expe}(T_i)- A_\chi T_i \chi_l^{\rm theo}(T_i) -B_\chi T_i\right)^2,
\end{equation}
where $T_{\min}$ is the lowest temperature where at least 6 pade approximants of the HTSE of $T\chi_l^{\rm theo}$ differ by less than 0.001 ($T_{\min}$ is evaluated for each set $\{J_\alpha\}$). $B_\chi$ is essential to account for some residual constant contribution and a small value less than $10^{-4}$ is expected.

The specific heat has a spin contribution and a phonon or lattice contribution. The phonon contribution is represented here by 3 modes (more do not help):
\begin{eqnarray}
	C_\mrm{p}^{\rm phonon}(T) &=& \sum_{k=1}^3 W_k P\left(\frac{T_{D,k}}T\right),\\
	P(x)&=& 9\left(\frac43 D_3(x)-\frac{x}{\exp(x)-1}\right),\\
	D_3(x) &=& {\frac{3}{x^3}} \int_0^x dt {t^3 \over \mrm{e}^t - 1}.
\label{eq:phononModel}
\end{eqnarray}
The error on $C_\mrm{p}$ of the model is measured by $E_C$:
\begin{eqnarray}
	E_C=\sum_{T_i>T_{\min}} \left(C_\mrm{p}^{\rm expe} - A_C C_\mrm{p}^{\rm spin} - C_\mrm{p}^{\rm phonon}(T) \right)^2
	\label{eq:CpModel},
\end{eqnarray}
where $T_{\min}$ is the the lowest temperature where at least 6 pade approximants of the HTSE of $C_\mrm{p}^{\rm spin}$ differ by less than 0.001. $A_C$ and $W_k$ are linear parameters, and $T_{D,k}$ non linear ones.

The set $\{J_\alpha\}$ of exchange energies varies on a grid of step $0.25\, K$, running from $-50\,K$ to $50\,K$.
For each $\{J_\alpha\}$, $\chi_l^{\rm theo}$ and $C_\mrm{p}^{\rm spin}$ are evaluated at the temperature of experiments. Then $A_\chi$, $B_\chi$, $A_C$, $W_k$,  and $T_{D,k}$ are caculated by minimizing $E_\chi$ and $E_C$. The total error of such fit is $E_\chi+E_C$. The best fit minimizes this total error. We found that the phonon Debye temperatures stay close to 300 K and 600 K for the two lowest temperatures whereas the last temperature is much higher around 3000 K. The respective weights $W_k$ are close to 28, 38 and 120.

\begin{figure*}[h]
  \begin{center}
\includegraphics[width=0.96\textwidth]{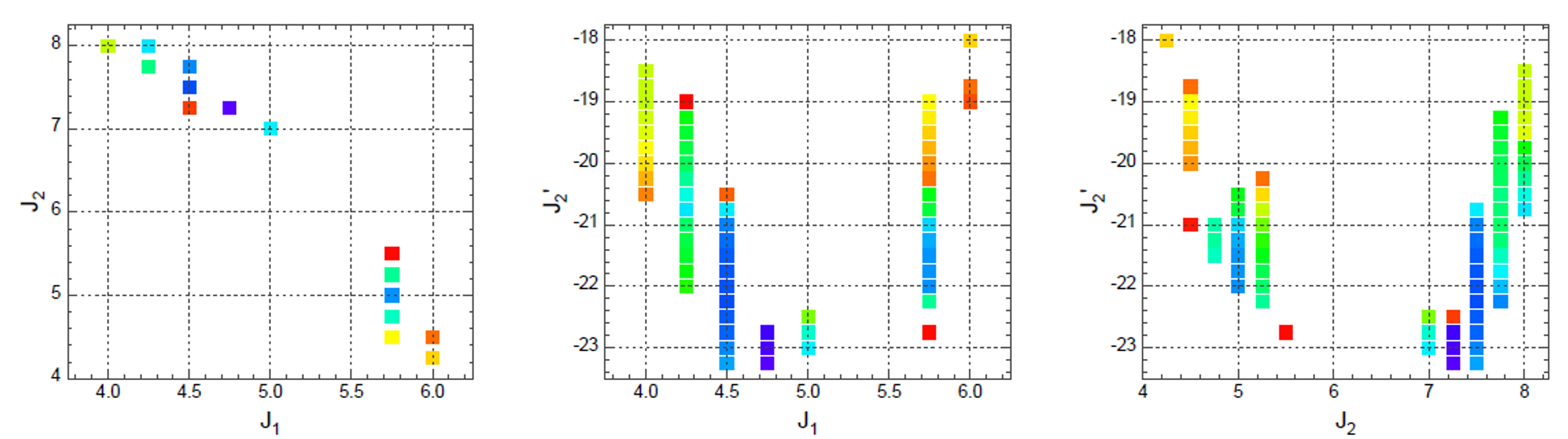 }
     \caption{Sets of parameters to represent both the specific heat and the magnetic susceptibility.
     The colors reflects the amplitude of the error, from the minimum error in violet to twice this minimum in red (using a rainbow color palette).
     }
    \label{fig:err}
  \end{center}
\end{figure*}

Fig.~\ref{fig:err} represents the best sets $\{J_1,\,J_2,\,J_2'\}$.
The points are selected such that $E=E_\chi/E_\chi^{\min}+E_C/E_C^{\min}<4$ while the minimum of this value is 2. The colors are scaled according to a rainbow palette with the color violet corresponding the value 2 and red for the value 4.
For $E>4$, the $\chi$-fit deviates significantly for experiment data.
When several sets fall on the same points on the pictures, only the best one is visible.
We see on Fig.~\ref{fig:err} that $J_1$ and $J_2$ are well correlated whereas $J_2'$ is more disperse.
A conservative set with error bar will be $\{J_1=5\pm1\,K,\ J_2=6\pm2\,K,\ J_2'=-21\pm3\, K\}$.
The best fits are found for $\{J_1=4.5\pm0.5\,K,\ J_2=7.5\pm0.5\,K,\ J_2'=-22\pm2\, K\}$.

\end{document}